\documentclass[12pt]{article}
\usepackage{amssymb}
\begin{document}
\title{
\[ \vspace{-7cm} \]
\noindent\hfill\hbox{\rm \large Alberta Thy 09-03} \vspace {5cm}\\
A Monte Carlo Test of\\ the Optimal Jet Definition\footnote{
Talk given by E. Jankowski at Lake Louise Winter Institute:
\emph{Particles and the Universe}, Lake Louise, Canada, February 16-22, 2003}}
\author{D.\ Yu.\ Grigoriev$^{1,2,\mathrm{a}}$,
E.\ Jankowski$^{3,\mathrm{b}}$,
F. V. Tkachov$^{2,\mathrm{c}}$\\ \\
\normalsize $^1$Mathematical Physics, Natl.\ Univ.\ of Ireland Maynooth,\\
\normalsize Maynooth, Co.\ Kildare, Ireland\\
\normalsize $^2$Institute for Nuclear Research of RAS,\\
\normalsize Moscow 117312, Russia\\
\normalsize $^3$Department of Physics, University of Alberta,\\
\normalsize Edmonton, AB, T6G 2J1, Canada\\
\normalsize $^\mathrm{a}$E-mail: dima@thphys.may.ie\\
\normalsize $^\mathrm{b}$E-mail: ejankows@phys.ualberta.ca\\
\normalsize $^\mathrm{c}$E-mail: ftkachov@ms2.inr.ac.ru}
\maketitle
\abstract{We summarize the Optimal Jet Definition and present the
result of a benchmark Monte Carlo test based on the $W$-boson mass
extraction from fully hadronic decays of pairs of $W$'s.}
\section{Introduction}
Jets of hadrons which appear in the final states of scattering experiments
in high energy physics correspond, to the first approximation,
to quarks and gluons produced in the collisions.
Quarks and gluons, interacting strongly,
are not observed as free particles. Only some combinations of them,
hadrons, can avoid the strong interaction at large distances and only those
combinations appear in experiments. If the energy of the colliding particles
is high enough, the quarks and gluons produced in the collision manifest
themselves as jets of hadrons which move roughly in the same direction as
the quarks and gluons originating them.

Let us consider an example high energy event.
An electron and positron collide at the CM energy equal
to 180 GeV. The electron and positron annihilate and a pair of $W$-bosons
is produced.
Each of the $W$'s decays into two quarks. When the quarks move away from
each other, potential energy of the strong interaction between them grows
quickly and new pairs of quarks and antiquarks are created out of this energy.
The many quarks and antiquarks combine into colorless hadrons which form
4 or more jets.

We are interested, for instance, in extracting the $W$-boson mass
from a collection of events similar to the one described above.
It would be much easier if we were able to observe directly the quarks
coming from decaying W's.
But we observe jets of hadrons instead and when we make the analysis
we have to deal with the jets. And this may not be always easy.
Jets may be wide and/or overlap. It is hard to say even how many jets we have
and how to share the particles between them.

Another aspect is that when we have a procedure to recognize and
reconstruct jets it may give different answers for the same physical
process depending whether it is applied at the level of quarks and gluons in theoretical calculations or at the level of hadrons from
Monte Carlo simulations or at the level of calorimeter cells
in experiments.

The Optimal Jet Definition avoids most of the problems of the conventional
schemes. The derivation of OJD from the properties of the strong interaction
and specifics of measurements involving multi-hadronic final states
is contained in \cite{tka-tjd}, \cite{tka-wij}.
A short introduction to the subject is \cite{letter}.
An efficient FORTRAN 77 implementation of OJD, called the Optimal Jet
Finder (OJF), is described in \cite{paper-ojf} and the source code is
available from \cite{source-ojf}. Below we summarize OJD and
present the result of a benchmark Monte Carlo test based
on the $W$-boson mass extraction from fully hadronic decays
of pairs of $W$'s.  
\section{Jet algorithms}
The analysis of events with many hadrons
is often performed with the use of so called jet algorithms.
A jet algorithm is a procedure to associate
the particles into jets. It decides which particle belongs to which jet.
Often it determines also how many jets there are.
(When we say particles it may mean as well calorimeter cells
or towers when the analysis is applied to experimental data or partons
in theoretical calculations.)

After the content of each jet is known, some rule is chosen
to compute the properties of the jet from the properties of the particles
that belong to that jet. A simple and logical prescription,
but not necessarily the only possible (see \cite{run2-jp} for discussion),
is that the \mbox{4-momentum} of the jet, $q_\mathrm{jet}$,
is the sum of 4-momenta
$p_a$ of all particles that belong to that jet:
$q_\mathrm{jet}=\sum _\mathrm{the\;jet} p_a.$

There have been many jet definitions developed by various
collaborations
over the years. Examples are the class of cone algorithms (various variants)
and the family of successive recombination algorithms such as $k_\mathrm{T}$
(Durham), Jade, Geneva. 

Cone algorithms define a jet as all particles within a cone of fixed radius.
The axis of the cone is found, for instance, from the requirement that
it coincides with the direction of the net 3-momentum of all particles within
the cone.

Successive recombination algorithms, in the simplest variant, work
as follows. The ``distance'' $d_{ab}$ between any two particles is computed
according to some definition, for example,
$d_{ab}^2=E_a E_b\left(1-\cos\theta_{ab}\right)$ for JADE
and $d_{ab}^2=\min\left(E_a^2,E_b^2\right)\left(1-\cos\theta_{ab}\right)$
for $k_\mathrm{T}$, where $E_a$ is the energy of the $a$-th particle
and $\theta_{ab}$ is the angle between the \mbox{$a$-th} and the $b$-th
particles.
Then the pair with the smallest difference is merged into one pseudo-particle
with the 4-momentum given (for example) by $p_{ab}=p_a+p_b$. In that way
the number of \mbox{(pseudo-)} particles is reduced by one. The procedure is
repeated until the required number of pseudo-particles is left
(if we know in advance how many jets we want) or until
$d_{ab}>y_\mathrm{cut}$ for all $a$, $b$, where $y_\mathrm{cut}$ is some
chosen parameter. The remaining pseudo-particles are the final jets.
The described scheme corresponds to so called binary algorithms as
they merge only two particles at a time ($2 \rightarrow 1$).
Other variants may correspond to $3 \rightarrow 2$ or more generally
to $m \rightarrow n$.

With many available jet definitions, an obvious question is
how to decide which algorithm should be used. It should be clear
that the jets are defined (through the jet algorithm used) for the purpose
of data analysis. In the example used it is the W-boson mass extraction.
In this case we can measure how good the jet definition is
based on how small the uncertainty in the extracted mass is.
On this idea we based our benchmark test of the Optimal Jet Definition.
\section{Optimal Jet Definition}
The OJD works as follows. It starts with a list of particles 
(hadrons, calorimeter cells, partons) and ends with a list of jets.
To find the final jet configuration we define $\Omega_R$,
some function of a jet configuration.
The momenta of the input particles enter $\Omega_R$ as parameters.
The final, optimal jet configuration is found
as the configuration that minimizes $\Omega_R$.

The essential feature of this jet definition is that it takes into
account the global structure of the energy flow of the event.
Above mentioned binary algorithms take at a time only two closest particles
into account, to decide whether to merge them or not.

A jet configuration is described by the so-called recombination matrix
$z_{aj}$, where $a$=1,2,...,$N_\mathrm{part}$ indexes the input particles
with 4-momenta $p_a$ and $j$=1,2,...,$N_\mathrm{jets}$ indexes the jets.
$z_{aj}$ is interpreted as the fraction of the $a$-th particle that goes into
formation of the $j$-th jet. The conventional schemes correspond to 
restricting $z_{aj}$ to either one or zero depending on whether or not
the $a$-th particle belongs to the $j$-th jet. Here we require only
that $0 \le z_{aj}  \le 1$ and $\sum_j z_{aj}  \le 1$.
The 4-momentum of the $j$-th jet is given by:
$q_j  = \sum_a z_{aj} p_a.$
The \mbox{4-direction} of the $j$-th jet is defined as
$\tilde{q}_j=\left(1,\mathbf{\hat{q}}_j\right)$,
where $\mathbf{\hat{q}}_j=\mathbf{q}_j/\left|\mathbf{q}_j\right|$
is the unit direction vector obtained from
$q_j=\left(E_j,\mathbf{q}_j\right)$.
The explicit form of $\Omega_R$ is:
$\Omega_R = 
\frac{2}{R^2} \sum_j q_j \tilde{q}_j
+ \sum_a \left(1-\sum_j z_{aj}\right) E_a.$
The first term in the above equation ``measures'' the width of the jets
and the second is the fraction of the energy of the event
that does not take part in any jet formation. The positive parameter $R$ 
has the similar meaning to the radius parameter
in cone algorithms in the sense that a smaller value of $R$ results
in narrower jets and more energy left outside jets. A large $(\gtrsim2)$
value of $R$ forces the energy left outside jets to zero.   

If the number of jets that the event should be reconstructed to
is already known one finds $z_{aj}$ that minimizes $\Omega_R$ given in the
above equation. This value of $z_{aj}$ describes the final desired
configuration of jets.  
The minimization problem is non-trivial because of the large dimension
of the domain in which to search the global minimum,
$N_\mathrm{part} \times N_\mathrm{jets}=O\left(100\mathrm{-}1000\right)$
of continuous variables $z_{aj}.$
However, it is possible to solve it due to the known analytical structure
of $\Omega_R$ and the regular structure of the domain of $z_{aj}$.
An efficient implementation, called the Optimal Jet Finder
(OJF), is described in detail in \cite{paper-ojf} and the FORTRAN 77
code is available from \cite{source-ojf}.
The program starts with some initial value of $z_{aj}$,
which in the simplest case can be entirely random,
and descends iteratively into the local minimum of $\Omega_R$.
In order to find the global minimum, random initial values of $z_{aj}$
are generated a couple of times ($n_\mathrm{tries}$) and the deepest
minimum is chosen out of the local minima obtained at each try.

If the number of jets should be determined in the process of jet
finding, one repeats the above described reconstruction for the
number of jets equal to 1,2,3,... and takes the smallest number of jets
for which the minimum of $\Omega_R$ is sufficiently small, i.e.
$\Omega_R<\omega_\mathrm{cut}$, where $\omega_\mathrm{cut}$ is
a positive parameter chosen by the user. $\omega_\mathrm{cut}$ has
a similar meaning to the $y_\mathrm{cut}$ parameter in the successive
recombination algorithms.

The shapes of jets are determined dynamically in OJD (as opposed
to the fixed shapes of cones in the cone algorithms). Jet overlaps
are handled automatically without necessity of any arbitrary prescriptions.
OJD is independent of whether input particles are split into collinear groups
(collinear safety). OJD is also infrared safe, i.e. any soft particle
radiation results in soft (small) only change in the structure of jets.
(So, it avoids the serious problems of cone algorithms based on seeds.) 
OJD, as opposed to successive recombination algorithms, takes into account
the global structure of the energy flow in the event (rather then merging
a single pair of particles at a time).  
\section{Details of the test}
We performed a simple, benchmark Monte Carlo test of the Optimal 
Jet Definition.
The analysis was modeled after a similar one performed by
the OPAL collaboration from LEP II data \cite{mass-W-opal}.

We simulated the process
$e^+e^- \to W^+W^- \to \mathrm{hadrons}$
at CM energy of 180 GeV using PYTHIA 6.2 \cite{pythia}.
We reconstructed each event to 4-jets using OJF
and two binary jet algorithms: $k_\mathrm{T}$ and Jade for comparison.
For OJF, we chose $R\!=\!2$ and employed the most primitive variant
of OJF-based algorithm with a fixed $n_\mathrm{tries}$=10 for all events.
The jets can be combined into two pairs (supposedly resulting from decays
of the $W$'s) in three different ways.
We chose the combination with the smallest difference in invariant masses
between the two pairs and calculated the average $m$ of the two masses.
We generated the probability distribution $\pi_M (m)$ with the $W$-boson mass
$M$ as a parameter. The smallest error of parameter estimation
corresponding to the number $N_\mathrm{exp}$ of experimental events
(as given by Rao-Frechet-Cramer theorem) is
$\delta M_\mathrm{exp} \cong
\left[
N_\mathrm{exp}
\int \mathrm{d}m \left(
\partial \ln \pi_M \left(m\right)/\partial M
\right)^2
\right]^{-\frac{1}{2}}.$
We can use this number directly to evaluate the jet algorithms.
\section{Results}
The statistical error $\delta M_\mathrm{exp}$ of the $W$-boson mass
corresponding to 1000 experimental events is displayed in the table below:
\begin{center}
\begin{tabular}{|ccc|}
\hline
ALGORITHM &\phantom{100}& $\delta M_\mathrm{exp}\pm3\;\mathrm{MeV}$\\
\hline
OJD/OJF &\phantom{100}& 106\\
$k_\mathrm{T}$ &\phantom{100}& 105\\
JADE &\phantom{100}& 118\\
\hline
\end{tabular}\\
\end{center}
(The error of 3 MeV in our results is dominated by the uncertainties
in the numerical differentiation with respect to $M$.)
Within the obtained precision Durham and OJF are equivalent
with respect to the accuracy, JADE appears to be worse.

An important aspect is the speed of the algorithms.
The average processing time per event depends on the number of particles
or detector cells in the input $N_\mathrm{part}$.
We observed the following empirical relations (time in seconds):
$1.2 \times 10^{-8} \times N_\mathrm{part}^3$ for $k_\mathrm{T}$
and $1.0 \times 10^{-4} \times N_\mathrm{part}  \times n_\mathrm{tries}$
for OJF. $N_\mathrm{part}$ varied from 50 to 170 in our sample,
with the mean value of 83. However, the behavior was verified
for $N_\mathrm{part}$ up to 1700
by splitting each particle into 10 collinear fragments
(similarly to how a particle may hit several detector cells).

We observe that OJF is slower for small number of particles
or detector cells whereas for a large number of particles it
appears to be relatively much faster.
In the process we studied it starts to be more efficient for
$N_\mathrm{part} \approx 90\sqrt{n_\mathrm{tries}}$.

It may be a strong advantage. For instance \cite{run2-jp},
in the CDF or D0 data analysis, where binary $k_\mathrm{T}$
algorithm is commonly used, it is not possible to analyze data directly
from the calorimeter cells or even towers because it would take forever.
The preclustering procedure (defined separately from the jet algorithm)
is necessary to reduce input data to approximately 200 preclusters.
With OJF, it is possible to test how the preclustering step affects
the results or even skip it altogether.
\section{Summary}
We performed a Monte Carlo test of the Optimal Jet Definition.
We found that in the process we studied it gives the same accuracy
as the best algorithm applied previously to the similar analysis.
OJD offers new options yet to be explored, 
e.g. the weighting of events (according to the value of $\Omega$) 
to enhance the precision.
We found that the already available implementation of OJD 
is very time efficient for analyses at the level of calorimeter cells. \\ \\
\textbf{Acknowledgments.}
We are grateful to Andrzej Czarnecki and James Pinfold
for helpful discussions.

\end{document}